# Generalized Surface Polaritons and their quantum spin Hall effect

Yadong Xu, Jian-Hua Jiang* & Huanyang Chen[†]

*College of Physics, Optoelectronics and Energy & Collaborative Innovation Center of Suzhou Nano Science and Technology, Soochow University, No.1 Shizi Street, Suzhou 215006, China*

**Abstract:**

Surface polaritons, e.g., surface plasmon polaritons, are invaluable tools in nanophotonics. However, considerable plasmon loss narrows the application regime of plasmonic devices. Here we reveal some general conditions for lossless surface polaritons to emerge at the interface of a gain and a loss media. The gain medium does not only compensate the energy loss, but also modifies surface wave oscillation mechanisms. A new type of surface polaritons induced by the sign switch of the imaginary part of the permittivity across the interface is discovered. The surface polaritons exhibit spin Hall effect due to spin-momentum locking and unique Berry phase. The spin Hall coefficient changes the sign across the parity-time symmetric limit and becomes quantized for perfect metal-dielectric interface and for dielectric-dielectric interface with very large permittivity contrast, carrying opposite topological numbers. Our study opens a new direction for manipulating light with surface polaritons in non-Hermitian optical systems.

[*] jianhuajiang@suda.edu.cn
[†] chy@suda.edu.cn

**Introductions**

Surface plasmon polaritons (SPPs) have served as useful tools to a wide span of scientific research using nanophotonics [1, 2]. The SPPs are essentially electromagnetic (EM) waves that propagate along an interface between a metal and a dielectric medium, while exponentially decay away from it. Their remarkable features include strong concentration of EM fields at subwavelength scales, enabling applications in the frontiers of technology ranging from molding light flow at nanoscales [3], imaging beyond the diffraction limit [4] to biosensing [5]. Lately, it was shown that as a type of surface waves, the SPPs with strong spin-momentum locking, exhibit intrinsic quantum spin Hall effect (QSHE) of light, thereby bringing new concept to SPPs as well as to dielectric-metal structures that are analogy to topological insulators for electrons [6].

However, such applications are hindered greatly by the intrinsic metallic Ohmic loss which results in substantial attenuation of the SPPs travelling along the metal-dielectric interfaces [7]. For instance, for silver nanowire (silver has the lowest loss at visible frequencies among metals), the propagation length of excited SPPs cannot be larger than ten micrometers at a working wavelength of 785nm [8]. Currently one of the most promising solutions to this issue is by introducing optical gain to the dielectric medium adjacent to the metal [9, 10]. Partial compensation of plasmonic loss were demonstrated experimentally at visible frequencies, by using several types of gain medium, such as pumped dye solution [11, 12], dipolar gain medium[13], fluorescent polymer[14] and indium gallium nitride core[15]. Furthermore, ultra-high optical gain was realized for efficient loss compensation at visible spectrum, based on hybrid plasmonic waveguides [16]. There is no study reported so far on how much gain is needed for perfect compensation (i.e., the SPPs propagate without attenuation). Here we develop a general theory for the sufficient and necessary conditions of stable, lossless surface polaritons to emerge at the loss-gain interfaces. We find that the gain needed for compensation first increases with loss but then decreases and eventually vanishes for very strong loss. More importantly, beside the energy-compensated SPPs, stable and lossless surface polaritons also emerge at the interface of two dielectric materials induced by the sign changes of the imaginary part of the complex permittivity (i.e., a loss-gain interface). Surface polaritons at parity-time symmetric interfaces are special cases for loss-gain induced surface polaritons. Like SPPs, loss-gain induced surface polaritons also exhibits spin Hall effect and nontrivial Berry phase. Remarkably, the spin Hall effect becomes quantized for epsilon-near-zero (ENZ) materials with the quantum number opposite to that of the conventional SPPs for lossless metal-dielectric interfaces. These surprising findings provide new instrumentals for subwavelength nanophotonics, particularly for disciplines with demanding optical coherences.

**Results**

The simplest geometry supporting the well-known SPPs, is a single flat interface (as shown in Fig. 1a) between a dielectric material with permittivity $\varepsilon_1$ (for the half-space with z>0), and a metal with permittivity $\varepsilon_2$ (for the half-space with z<0). Such an interface breaks the dual symmetry between the electric and magnetic properties. Only transverse-magnetic (TM) polarization supports SPPs. By solving the Maxwell's equations together with boundary

conditions, the relevant dispersion relation can be obtained as

$$\beta(\omega)=k_0\sqrt{\frac{\varepsilon_1\varepsilon_2(\omega)}{\varepsilon_1+\varepsilon_2(\omega)}} \qquad (1)$$

where $\beta(\omega)$ is the propagating wave vector of SPPs, and $k_0=\omega/c$ is the wave vector in free space. The common knowledge for satisfying this condition, is that both permittivity $\varepsilon_1$ of dielectric and the frequency-dependent permittivity $\varepsilon_2(\omega)$ of metal must have opposite signs and $-\varepsilon_2>\varepsilon_1$. For real metals with loss, the frequency-dependent permittivity $\varepsilon_2(\omega)$ is not a purely real value, but a complex number. The equation (1) is still valid for a complex $\varepsilon_2(\omega)$, with its wave vector $\beta(\omega)$ of SPPs also complex. The propagation of SPPs is then damped with a decay length $l=1/\text{Im}[\beta(\omega)]$. Such damping is a major obstacle for the current SPP-based applications in plasmonics. Including gain in the dielectric material can help to reduce SPPs' loss.

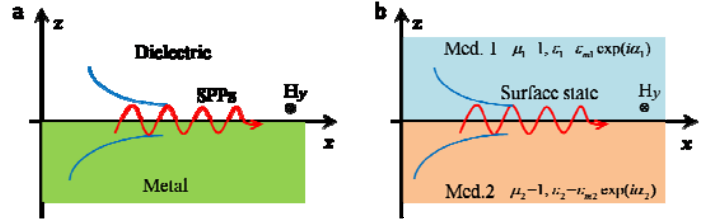

*Figure 1 | Surface polaritons existed at an interface of two media. **a**, Surface plasmon polaritons propagate along the interface between a dielectric (e.g. vacuum) and a metal, evanescently confined in the perpendicular direction. **b**, Generalized surface polaritons are supported by the interface between nonmagnetic response medium 1 with permittivity $\varepsilon_1=\varepsilon_{m1}\exp(i\alpha_1)$ and medium 2 with permittivity $\varepsilon_2=\varepsilon_{m2}\exp(i\alpha_2)$. These surface states are of TM polarizations.*

Starting from the general situation, as shown in Fig. 1b, we assume that medium 1 is nonmagnetic with permittivity $\varepsilon_1=\varepsilon_{m1}\exp(i\alpha_1)$, and the nonmagnetic medium 2 with a permittivity $\varepsilon_2=\varepsilon_{m2}\exp(i\alpha_2)$. We call $\alpha_1$ and $\alpha_2$ the permittivity phases. Such an interface supports only surface polaritons of TM polarization. Equation (1) is still valid for complex permittivity. We then obtain the following results directly,

$$\beta(\omega)=k_0\sqrt{\frac{\varepsilon_{m1}\varepsilon_{m2}}{\Im-i\xi}} \quad, \qquad (2)$$

where $\Im=\varepsilon_{m1}\cos(\alpha_2)+\varepsilon_{m2}\cos(\alpha_1)$ and $\xi=\varepsilon_{m1}\sin(\alpha_2)+\varepsilon_{m2}\sin(\alpha_1)$. To achieve a purely real wave vector, the following conditions must be satisfied,

$$\varepsilon_{m1}\sin(\alpha_2)+\varepsilon_{m2}\sin(\alpha_1)=0, \qquad (3a)$$

$$\varepsilon_{m1}\cos(\alpha_2)+\varepsilon_{m2}\cos(\alpha_1)\geq 0. \qquad (3b)$$

The criterion for lossless surface polaritons can be discussed with the awareness of the scale invariance of the Maxwell's equation. That is, by rescaling the frequency one can always set the absolute value of the relative permittivity of medium 1 as unity ($\varepsilon_{m1}=1$). After that, the complex permittivity $\varepsilon_2$ is a function of the phase $\alpha_1$ using equation (3a) (or reversely, $\alpha_1$ is a function of the complex permittivity $\varepsilon_2$). We focus on the situation where medium 1 is a dielectric with gain, i.e., $-\pi/2 \leq \alpha_1 < 0$, and medium 2 is a lossy material (can be metal or dielectric). Solving equation (3a) with condition (3b), we plot the $\varepsilon$-diagram for the required medium 2 with different $\alpha_1$, as shown by Fig. 2a. The dashed unit circle represents the complex permittivity $\varepsilon_1$. All of the curves (denote the relationship of the real part and the imaginary part of the permittivity of medium 2, we would call them "relationship circles") are parts of some circles. Each circle, with diameter $1/|\sin(\alpha_1)|$, starts from the origin and ends at its point of intersection with the unit circle in the left.

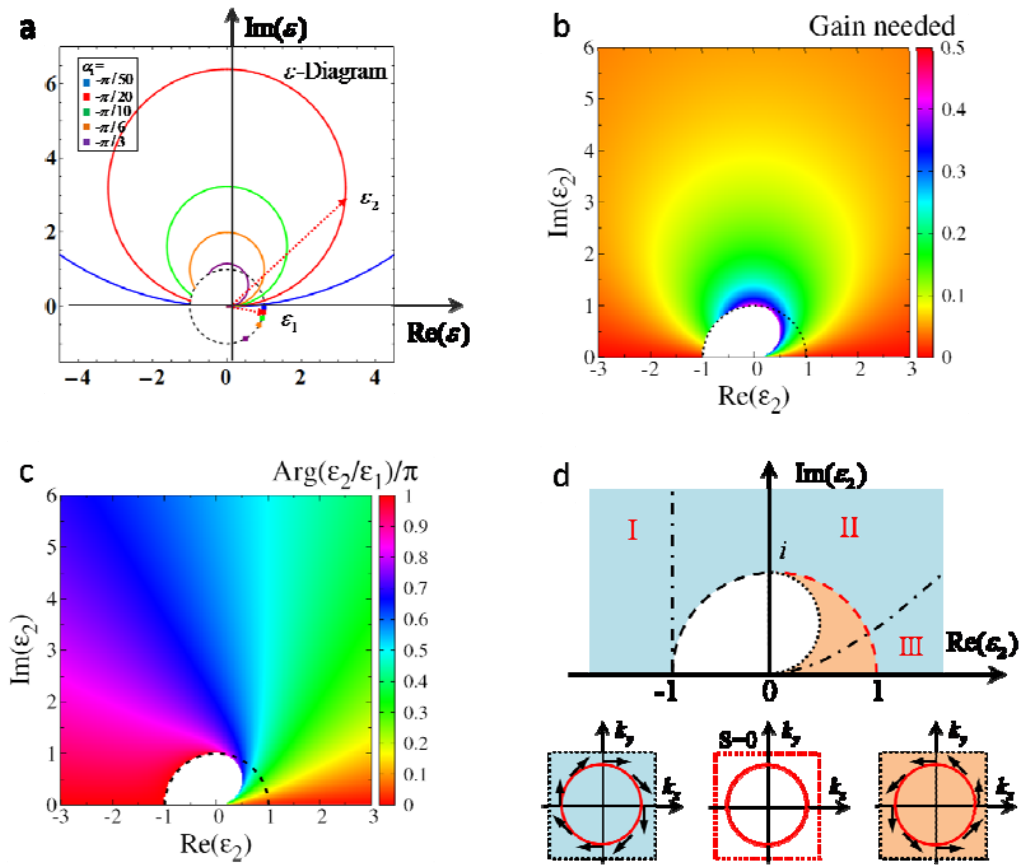

**Figure 2 | $\varepsilon$-Diagrams for emergence of lossless surface polaritons.** *The horizontal and vertical axis's correspond to the real and imaginary parts of the permittivity of medium 2, respectively. The dashed circle represents nonmagnetic materials with $|\varepsilon|=1$. **a**, For a given $\alpha_1$, the complex permittivity of medium 2 that supports lossless surface polaritons is given by a relationship circle. We plot five circles for five different values of $\alpha_1$. The relationship circles end at their intersection with the unit circle in the left half plane. **b**, The gain needed*

$|\sin(\alpha_1)|$ to support lossless surface polaritons as a function of the complex permittivity $\varepsilon_2$ for the whole plane of $\text{Re}[\varepsilon_2]$ and $\text{Im}[\varepsilon_2]$ (for lossy medium 2). The white region does not support lossless surface polaritons. **c,** Relative phase between the two complex permittivity of the two media for conditions with lossless polaritons. **d,** Phase diagram for the lossless surface polariton. The chained curves separate the three regions. The dashed curve represents the unit circle. The right-upper (red) quarter of the unit circle stands for the parity-time symmetric interfaces. Lower panel (left to right): (1) spin rotation on the iso-frequency contour of the surface polaritons in the regions outside the unit circle, (2) spin vanishes for the parity-time symmetric interfaces, (3) spin rotation in the regions inside the unit circle.

This diagram reveals precisely a pair of complex permittivity for the two media $\{\varepsilon_1=\exp(i\alpha_1), \varepsilon_2=\varepsilon_{m2}\exp(i\alpha_2)\}$, with $\varepsilon_{m2}=-\sin(\alpha_2)/\sin(\alpha_1)$ which support lossless surface polaritons at their interface. Surprisingly, lossless surface polaritons exist in a wide range of parameters, not only for $\text{Re}[\varepsilon_2]<-1$ as commonly believed. Particularly, it supports surface polaritons for materials with near zero or even positive $\text{Re}[\varepsilon_2]$. We also plot the gain needed $|\sin(\alpha_1)|$ to support the surface polaritons for different complex permittivity of medium 2 in Fig. 2b. Counter-intuitively, the gain needed becomes rather small when the loss of medium 2 is large. As also depicted by the relationship circles in Fig. 2a, the gain needed eventually becomes vanishingly small for very large loss in medium 2. For medium 2 with near zero $\text{Re}[\varepsilon_2]$, the required gain decreases monotonically with the loss of medium 2. In comparison, for the regime with $\text{Re}[\varepsilon_2]<-1$ (or $\text{Re}[\varepsilon_2]>0.5$) the gain needed first increases but then decreases as the loss of medium 2 increases. Practically achievable loss-gain balance can be realized in the regimes with either large loss or with very large absolute value of $\text{Re}[\varepsilon_2]$. For instance, the permittivity of silver is $\varepsilon_{Ag}=-27.48+0.3145i$ at a wavelength of 756nm [17], which can be compensated perfectly by a gain medium with complex permittivity $\varepsilon_{m1}=1-0.0004165i$.

To reveal the underlying mechanism, we recall that the necessary physical condition for the existence of surface polaritons is that the polarization along z direction changes the sign across the interface. This is essentially triggered by the sign change of the permittivity. In Fig. 2c we plot the relative phase between the complex permittivity of the two media when their interface supports lossless surface polaritons. The permittivity sign change mechanism for conventional SPPs are faithfully demonstrated as the relative permittivity phase $\pi$ along the horizontal axis for $\text{Re}[\varepsilon_2]<-1$. Besides, the relative permittivity phase $\pi$ is also realized at the left-upper quarter of the unit circle, which clearly indicates the mechanism of the loss-gain induced surface polaritons: They are caused by polarization concentration induced by the sign change of the imaginary part of the permittivity. The discontinuity of the polarization at the interface in fact exists also for other relative permittivity phases as manifested in Fig. 2c. The relative permittivity phase goes to zero when approaching the positive sector of the axis of $\text{Re}[\varepsilon_2]$, where the surface polaritons become unconfined (i.e., they become propagating waves). In general the confinement along $z$ direction weakens when the relative permittivity phase decreases.

The phase diagram of lossless surface polaritons is summarized in Fig. 2d. Region I represents conventional SPPs compensated with gain medium. Region II stands for the loss-gain induced surface polaritons discovered in this work. In region III the confinement of the surface polaritons in $z$ direction is weak and only surface-like waves are supported. The right-upper quarter of the unit circle (represented by the dashed curve) stands for the parity-time symmetric interfaces with $\varepsilon_2 = \varepsilon_1^*$. This only appears for interfaces between two dielectrics, i.e., $\text{Re}[\varepsilon_2] > 0$.

In the past few years much attention has been paid to the parity-time symmetry systems with numerous novel phenomena in optics [18,19,20] and associated devices emerged, such as unidirectional invisibility [21], lossless Talbot effect [22] and coherent perfect absorption [23, 24]. Our work thereby proposes a new effect for optical parity-time symmetry systems. We show below that the topology of the surface polariton, i.e., spin Hall coefficient (and the Berry phase), changes the sign upon crossing the parity-time symmetric line.

As for the stability of the lossless surface polaritons, for regions outside of the unit circle, the surface polaritons increases during propagation when the gain medium overcompensates, while it damps if the gain falls short. For the white region in the phase diagram, all surface polaritons are damped. The relationship circle also gives the information for the stability of the surface polaritons: outside of the circle surface polaritons increases during propagation, whereas inside the circle it damps (see Supplementary Materials Part I). There is no singularities in the dependence of the damping/gain of the surface polaritons on the small deviation of gain in the medium 2, unless for some special limit discussed below. Therefore, the lossless surface polaritons found in this work are stable.

Quite interestingly, we find that the topological properties of the surface polaritons are distinct at the two sides of the unit circle. Within the unit circle, the photonic spin winds counter-clockwise on the iso-frequency contour. Outside the unit circle, the spin winds clockwise. The winding number inside and outside the unit circle is hence +/-1, corresponding to different chiralities. The parity-time symmetric limit, i.e., the right-upper quarter of the unit circle, has zero winding number, which is consistent with the fact that U(1) Berry-phase vanishes in systems with parity-time symmetry.

There are two special limits in the phase diagram deserve to be mentioned: (1) The wave vector of the surface polariton goes infinite when approaching the left-upper quarter of the unit circle, where surface polaritons become unstable (see Supplementary Materials Part I). (2) The intersection between the vertical axis and the unit circle, $\varepsilon_2 = \varepsilon_1^* = i$, is a special limit that the surface polaritons emerge purely due to loss-gain discontinuity at the interface.

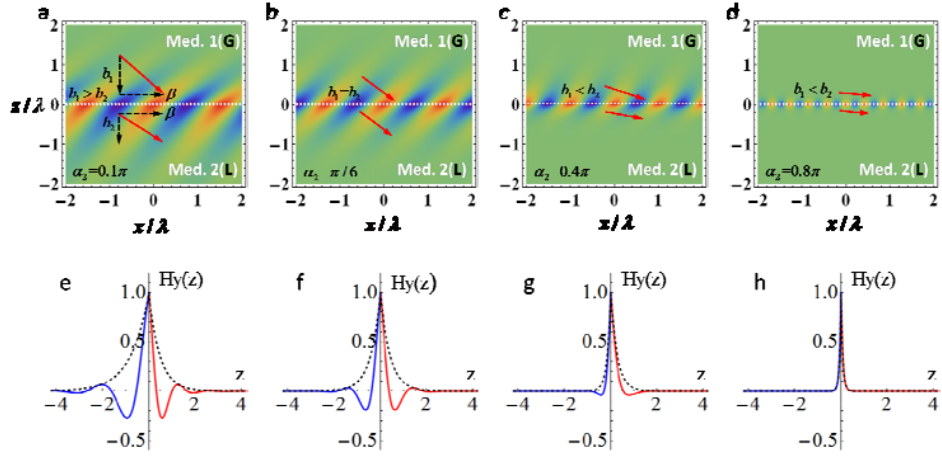

**Figure 3 | Analytical $H_y$ field patterns and distributions of lossless surface polaritons travelling along the interface between medium 1 with** $\varepsilon_1 = \exp(-\pi/6)$ **and medium 2 with** $\varepsilon_2 = \eta\exp(i\alpha_2)$. **a-d show to the field patterns for** $\alpha_2 = \pi/10$, $\pi/6$, $0.4\pi$ **and** $0.8\pi$, **respectively. e-h are corresponding field distributions in z direction.**

To explore in depth the physics of the lossless surface polaritons confined at the interface between gain and loss media, the field distributions are analyzed based on Maxwell's equations. The magnetic field of the lossless surface polaritons is expressed as,

$$H_y^1 = A\exp(i\beta x - \kappa_1 z), \quad \text{for } z>0, \tag{4a}$$

$$H_y^2 = A\exp(i\beta x + \kappa_2 z), \quad \text{for } z<0, \tag{4b}$$

where A is the amplitude coefficient, and $\kappa_{1(2)} = \sqrt{\beta^2 - \varepsilon_{1(2)}k_0^2}$. For demonstration, Fig. 3a-d show the field patterns of the lossless surface polaritons for a given $\alpha_1 = -\pi/6$ for different $\alpha_2 = \pi/10$, $\pi/6$, $2\pi/5$, and $4\pi/5$. In the figures the patterns of the magnetic field incline with respect to the interface, in particular when $\alpha_2$ is not large. Considering that $\kappa_{1(2)}$ are also complex, they can be expressed further by $\kappa_1 = a_1 + ib_1$ for gain medium 1, and $\kappa_2 = a_2 - ib_2$ for loss medium 2. Plugging these into equation (4), we get

$$H_y^1 = A\exp(i\beta x - ib_1 z)\exp(-a_1 z), \quad \text{for } z>0, \tag{5a}$$

$$H_y^2 = A\exp(i\beta x - ib_2 z)\exp(a_2 z), \quad \text{for } z<0, \tag{5b}$$

These equations give that: The wave vector of the lossless surface polaritons in gain medium is $\vec{\beta}_1 = \beta\hat{x} - b_1\hat{z}$, while in loss medium, $\vec{\beta}_2 = \beta\hat{x} - b_2\hat{z}$, as shown by the inset in Fig. 3a. This is the reason why the lossless surface polaritons incline with respect to the interface, which is quite different from the conventional SPPs. In the z direction, however, the field is not purely exponential-decay away from the interface, but rather becomes damped oscillations. When $\alpha_2$ is not large, the phenomenon of oscillation damps. While as $\alpha_2$ goes to $\pi+\alpha_1$, the damped oscillation becomes exponential decay. The damped oscillation indicates energy flow from the gain to the loss since the wave vectors along the $z$ direction for the oscillation, $b_1$ and $b_2$ has the same direction for both media.

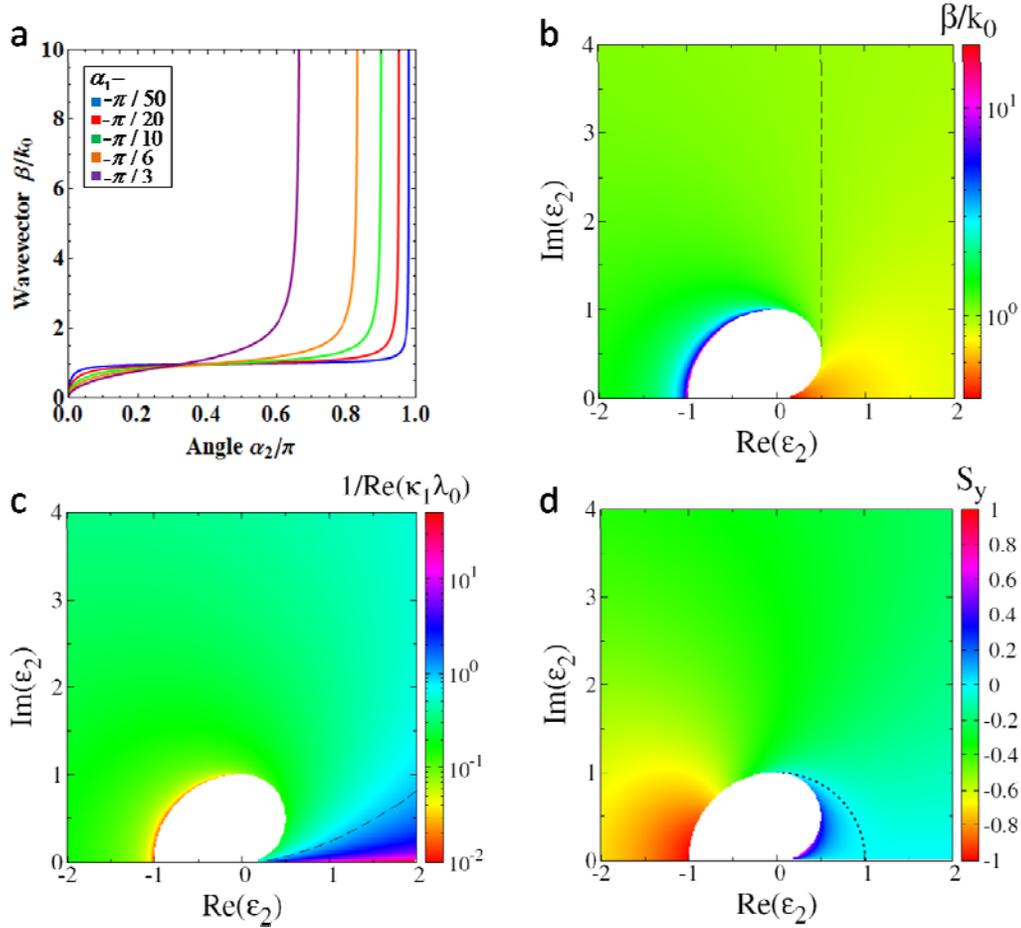

**Figure 4 | Properties of the surface polaritons. a,** The relationship between the wave vectors of the lossless surface polariton $\beta$ and $\alpha_2$, for fixed $\alpha_1$. The blue, red, green, orange and purple curves correspond to $\alpha_1 = -\pi/50, -\pi/20, -\pi/10, -\pi/6$ and $-\pi/3$, respectively. **b,** Full plot for the wave vector of the lossless surface polaritons as a function of the complex permittivity $\varepsilon_2$. The dashed curve denotes $\beta=k_0$. **c,** The normalized decay factor is defined as $\gamma = 1/\text{Re}[\kappa_1 \lambda_0]$, where $\lambda_0 = 2\pi/k_0$ is the working wavelength. The dashed curve stands for $\gamma=1$. **d,** The spin polarization of the surface polaritons.

The physical properties of the lossless surface polaritons are explored in more details in Fig. 4. We first show that the propagating wave vector of the surface polariton varies significantly for different $\alpha_2$. For $\pi/2 < \alpha_2 < \pi$, the propagating wave vector $\beta$ is very large. The wave vector goes to infinity in the limit with $|\varepsilon_2| \to 1$ if $\text{Re}[\varepsilon_2] \leq 0$. The lossless surface polaritons become unstable and disappear in this limit. The physics here is similar to that for SPPs at the metal-dielectric interface. For $0 < \alpha_2 < \pi/2$, the wave vector $\beta$ can be smaller than the wave vector $k_0$ in vacuum, and in the limit $\alpha_2 \to 0$ the wave vector $\beta$ vanishes. $\alpha_2 \to 0$ is hence the unconfined limit which is consistent with the total reflection at the dielectric-dielectric interface. A systematic survey of the wave vector of surface polaritons as a function of the complex permittivity $\varepsilon_2$ is shown in Fig. 4b. The dashed line denotes the case with $\beta=k_0$. Interestingly, the line lies on the condition $\text{Re}[\varepsilon_2]=0.5$ with $\text{Im}[\varepsilon_2] \geq 0.5$.

We also plot the normalized decay factor $\gamma=1/\text{Re}[\kappa_1 \lambda_0]$, as a function of the complex permittivity $\varepsilon_2$ in Fig. 4c. The dashed curve indicating $\gamma=1$ is in fact the line separating regions II and III in Fig. 2d. The confinement of the surface polariton is reduced in region III compared to region II, as shown in Fig. 4c. In region III the surface polaritons become surface-like waves. In regions I and II the confinement is strong enough to support subwavelength light focusing and imaging, for instance. In practice the thickness of the gain medium is not infinite but only several times of the decay length $1/\text{Re}[\kappa_1]$. Hence the strong confinement also enables applications of the surface polaritons with only a thin film of gain medium.

The spin polarization of the surface polaritons is studied in Fig. 4d. In the energy-compensated SPP region (i.e., region I in Fig. 2d), the spin polarization is negative. It becomes -1 in the limit of $\varepsilon_2 \to -1$. Away from this special point the spin polarization decreases. This spin polarization has been found recently as a topological property of the SPPs: the spin polarization is always perpendicular to the propagating wave vector and winds counter-clockwise. The integer winding of the spin polarization at an iso-frequency circle of the SPP spectrum gives rise to the unique topological properties. The spin-momentum locking and Berry phase can be exploited for, e.g., polarization control of light propagation. Here we find that, interestingly, in region II and region III, the spin polarization is highly tunable. In the limit of parity-time symmetry, i.e., the right-upper quarter of the unit circle (dashed curve) in Fig. 4d, the spin polarization vanishes. Moreover, outside the unit circle, the spin polarization is negative, while inside the unit circle the spin polarization is positive. The spin polarization becomes 1 at the boundary of the curve with $|\varepsilon_2|=\sin(\alpha_2)$ when $\alpha_2 \to 0$. Therefore, the topology of the surface polaritons is distinct from the ideal metal-dielectric interface, leading to opposite topological numbers (see Supplementary Materials Part II). The ability of manipulating the spin polarization via loss and gain will introduce new physics in surface polaritons and new methods of molding light flow at the interface via, for example, Berry-phases.

Note that if medium 1 is metallic gain medium with $-\pi \leq \alpha_1 < -\pi/2$, the permittivity of medium 2 can also be found in the $\varepsilon$- diagram, where stable lossless surface polaritons are supported at the interface (see Supplementary Materials Part III). However, for the issue of compensating the loss of metal, there seems to be no point to study this situation. Note that we have scaled $|\varepsilon_1|$ to unit. Our theory also applies to situations where both media are epsilon-near-zero metamaterials (or other interesting situations). Hence our work indicates that at interfaces between the epsilon-near-zero metamaterials with gain and loss, a stable lossless surface state can emerge [25].

**Discussions**

In summary, we have shown that the interface of loss medium and gain medium can support lossless surface polaritons. Many intriguing properties of such surface polaritons are uncovered. The conventional SPP mechanism is recovered for ideal metal-dielectric interface, while new mechanism and properties are discovered. Our work offers a more general theoretical framework for the study of surface polaritons and plasmonics, meanwhile opens a new route to exploit gain medium for surface polaritons with nontrivial Berry phase and topology.

**Acknowledgements**

This work is supported by the Postdoctoral Science Foundation of China (grant no. 2015M580456), the National Science Foundation of China for Excellent Young Scientists (grant no. 61322504), the Foundation for the Author of National Excellent Doctoral Dissertation of China (grant no. 201217), and the Priority Academic Program Development (PAPD) of Jiangsu Higher Education Institutions. J. H. J. acknowledges support from the start-up funding of Soochow University. Y. X. and H. C. conceived this idea. Y. X., H. C. and J. H. J. did the theoretical calculations and analyze all results. All the authors wrote the manuscript.